\documentclass[sigconf,xcolor=dvipsnames]{acmart}

\AtBeginDocument{%
  }

\copyrightyear{2026}
\acmYear{2026}
\setcopyright{cc}
\setcctype{by}
\acmConference[CHI '26]{Proceedings of the 2026 CHI Conference on Human Factors in Computing Systems}{April 13--17, 2026}{Barcelona, Spain}
\acmBooktitle{Proceedings of the 2026 CHI Conference on Human Factors in Computing Systems (CHI '26), April 13--17, 2026, Barcelona, Spain}
\acmPrice{}
\acmDOI{10.1145/3772318.3791534}
\acmISBN{979-8-4007-2278-3/2026/04}

\usepackage{xspace}
\newcommand{\tool}{\textit{TableTale}\xspace}

\newcommand{\revised}[1]{\textcolor{black}{#1}}
\newcommand{\smallpar}[1]{{\small (#1)}}

\newcommand{\m}[1]{\raisebox{0pt}[0.8\height][0.3\depth]{\colorbox{gray!20}{#1}}}
\usepackage{array}
\usepackage{enumitem}

\newcommand{\point}[1]{%
  \vspace{0.5em} 
  \noindent 
  \textbf{$\diamond$ \textit{#1}} 
}

\usepackage{multirow}
\newcommand{\semantic}[1]{\textcolor{teal}{\textbf{\textit{#1}}}}
\newcommand{\numeric}[1]{\textcolor{orange}{\textbf{\textit{#1}}}}
\newcommand{\structural}[1]{\textcolor{violet}{\textbf{\textit{#1}}}}

\usepackage{tikz}
\usetikzlibrary{positioning}  

\acmSubmissionID{2052}

\begin{document}

\title{TableTale: Reviving the Narrative Interplay Between Data Tables and Text in Scientific Papers
}


\author{Liangwei Wang}
\orcid{0000-0003-3481-3993}
\affiliation{
  \institution{The Hong Kong University of Science and Technology (Guangzhou)}
  \country{Guangzhou, China}
}
\email{lwang344@connect.hkust-gz.edu.cn}

\author{Zhengxuan Zhang}
\orcid{0000-0002-3370-1976}
\affiliation{
  \institution{The Hong Kong University of Science and Technology (Guangzhou)}
  \country{Guangzhou, China}
}
\email{zzhang393@connect.hkust-gz.edu.cn}

\author{Yifan Cao}
\orcid{0000-0002-5892-5052}
\affiliation{
  \institution{The Hong Kong University of Science and Technology}
  \country{Hong Kong SAR, China}
}
\email{caoyifan@ust.hk}

\author{Fugee Tsung}
\orcid{0000-0002-0575-8254}
\affiliation{
  \institution{The Hong Kong University of Science and Technology}
  \country{Hong Kong SAR, China}
}
\affiliation{
  \institution{The Hong Kong University of Science and Technology (Guangzhou)}
  \country{Guangzhou, China}
}
\email{season@ust.hk}

\author{Yuyu Luo}
\orcid{0000-0001-9530-3327}
\authornote{Yuyu Luo is the corresponding author.}
\affiliation{
  \institution{The Hong Kong University of Science and Technology (Guangzhou)}
  \country{Guangzhou, China}
}
\email{yuyuluo@hkust-gz.edu.cn}

\renewcommand{\shortauthors}{Wang et al.}

\begin{abstract}
Data tables play a central role in scientific papers. 
However, their meaning is often co-constructed with surrounding text through narrative interplay, making comprehension cognitively demanding for readers.
In this work, we explore how interfaces can better support this reading process. 
We conducted a formative study that revealed key characteristics of text-table narrative interplay, including linking mechanisms, multi-granularity alignments, and mention typologies, as well as a layered framework of readers’ intents. 
Informed by these insights, we present \tool, an augmented reading interface that enriches text with data tables at multiple granularities, including paragraphs, sentences, and mentions. 
\tool automatically constructs a document-level linking schema within the paper and progressively renders cascade visual cues on text and tables that unfold as readers move through the text. 
A within-subject study with 24 participants showed that \tool reduced cognitive workload and improved reading efficiency, demonstrating its potential to enhance paper reading and inform future reading interface design.
\end{abstract}



\begin{teaserfigure}
  \centering
  \includegraphics[width=0.94\linewidth]{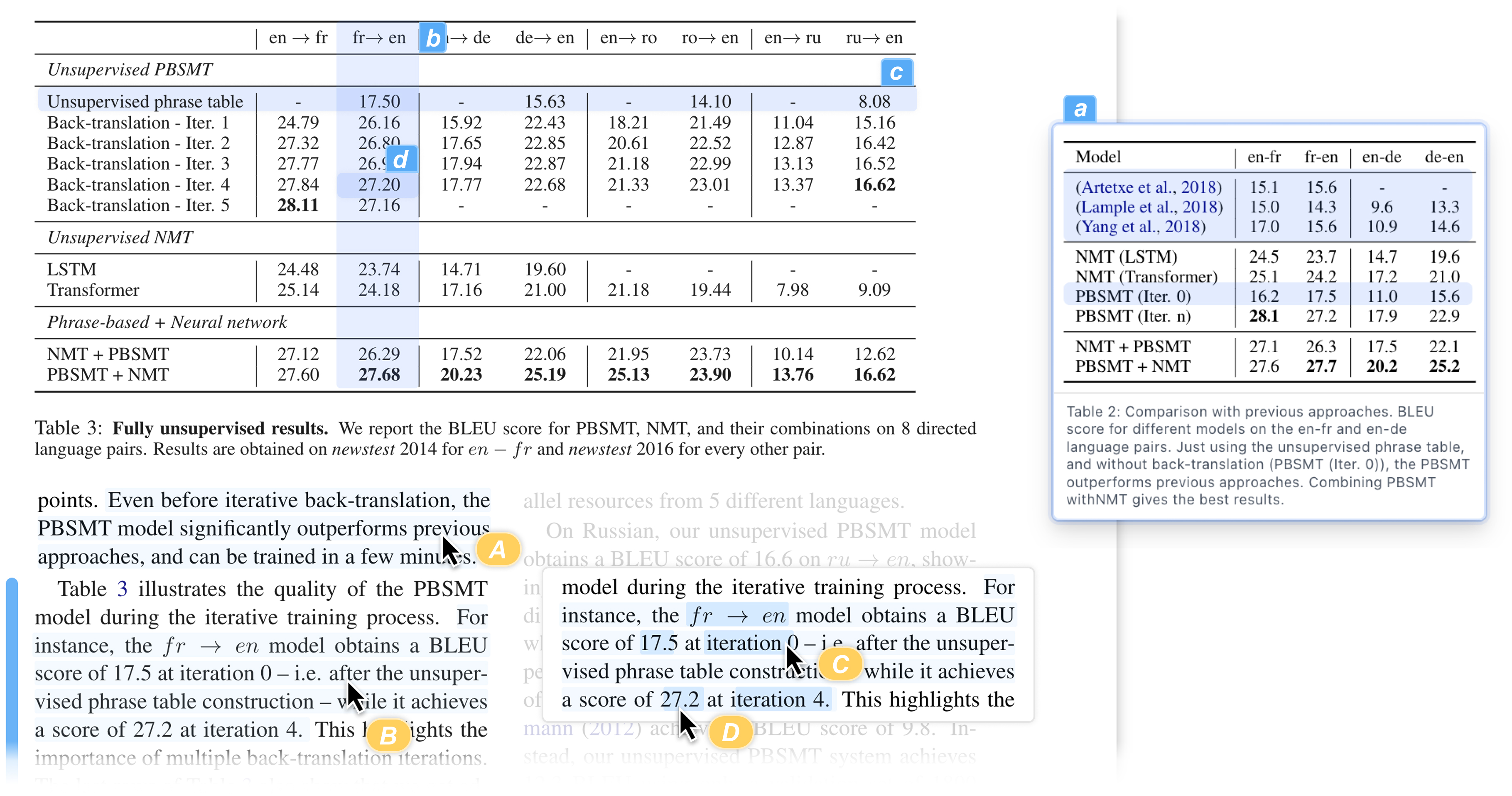}
  \caption{
 \textit{TableTale} is an interactive reading interface that reconstructs text-table interplay with cascade visual cues. 
 When the reader hovers over a sentence related to a table, the system shows sentence-level highlights, either on a mirror table when the table is off-screen (A-a) or as in-situ highlights when the table is visible (B-b). 
 Clicking on the sentence further reveals mention-level details (C-c, D-d), highlighting specific table regions, columns, rows, or cells.
  }
\Description{An annotated screenshot of the TableTale interface displaying a scientific paper on machine translation. The view consists of a main data table (Table 3) at the top and body text at the bottom. Four pairs of annotations (yellow circular cursors labeled A-D and blue rectangular highlights labeled a-d) demonstrate different interaction granularities. 
  1. Pair A-a: The cursor (A) hovers over a general sentence about model performance. This triggers a pop-up "mirror table" (a) on the right, displaying "Table 2" to show referenced data without scrolling.
  2. Pair B-b: The cursor (B) highlights the phrase "fr to en" in the text. A corresponding blue vertical bar (b) highlights the "fr to en" column in the main Table 3.
  3. Pair C-c: The cursor (C) focuses on the text "iteration 0". Inside the pop-up table, the specific row for "PBSMT (Iter. 0)" is highlighted in blue (c).
  4. Pair D-d: The cursor (D) hovers over the numerical value "27.2" in the text. The specific cell containing the value "27.20" within the main table is highlighted in blue (d), visually linking the textual statistic to its tabular source.}
  \label{fig:teaser}
\end{teaserfigure}

\maketitle

\section{INTRODUCTION}


Reading scientific papers is essential for understanding technological advances, recognizing unresolved problems, and inspiring future research.
Yet the process remains cognitively demanding~\cite{mayer2005cognitive,o1997comparison}. 
Scholars must integrate complex information such as figures, tables, and formulas to fully understand the contributions presented in a paper.
Among these, data tables are particularly critical as they encapsulate key statistics, experimental results, and configurations. 
Research has shown that tables function as cognitive scaffolds, helping readers organize information and perform effective comparisons~\cite{bartram2022untidy}, and in many cases, convey much of the underlying narrative of a paper~\cite{inskip2017getting}.

\begin{figure*}[t!]
  \centering
  \includegraphics[width=\linewidth]{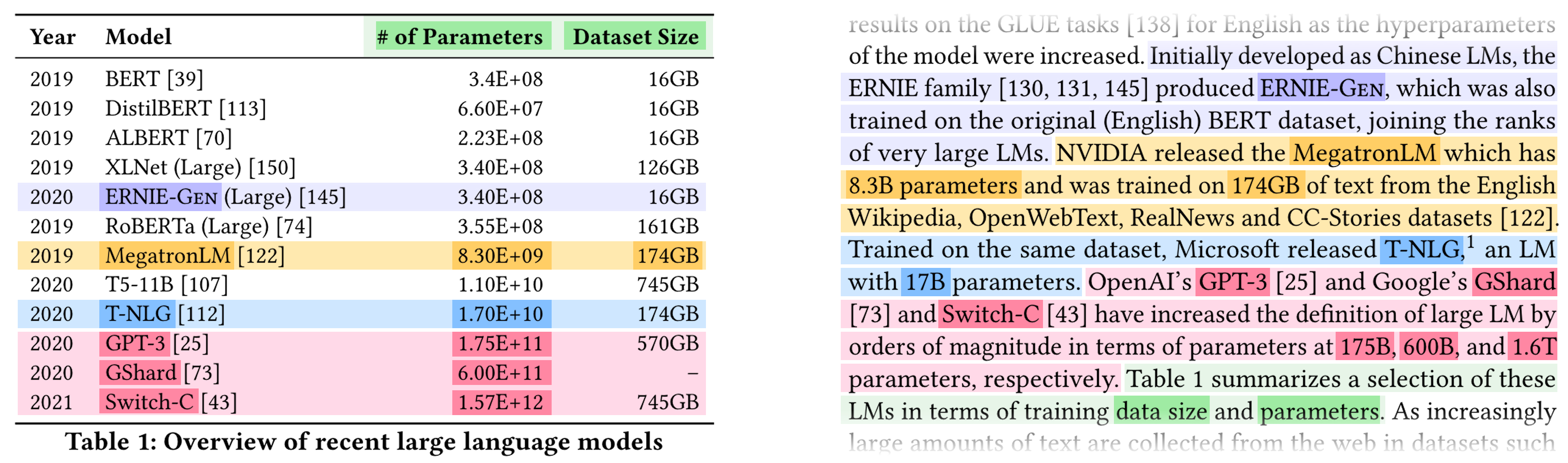}
  \caption{
  Example of narrative interplay between text and data table.
An excerpt from Bender et al.~\cite{bender2021dangers} shows how text and the related table jointly construct meaning. 
The narrative references the table through (1) semantic links using model names, such as MegatronLM and GPT-3;
(2) numeric links requiring readers to reconcile textual values (e.g., 1.6T parameters) with scientific notation in the table (e.g., 1.57E+12); 
and (3) structural links that reuse descriptors such as parameters and data size. 
}
\Description{A diagram illustrating specific semantic and numeric links between a data table and a text paragraph. The left side is a table titled Overview of recent large language models with columns for Year, Model, Number of Parameters, and Dataset Size. The right side is a text excerpt. Color-coded highlights indicate correspondences between the two. Green highlights the table headers Number of Parameters and Dataset Size, linking them to the text phrase data size and parameters. Yellow highlights the table row for MegatronLM (Year 2019, 8.30E+09 parameters, 174GB dataset), connecting it to text stating 8.3B parameters and 174GB of text. Blue highlights the rows for ERNIE-GEN (Year 2020) and T-NLG (Year 2020, 1.70E+10 parameters), linking to text mentions of ERNIE-GEN and T-NLG with 17B parameters. Pink highlights the rows for GPT-3 (1.75E+11 parameters), GShard (6.00E+11 parameters), and Switch-C (1.57E+12 parameters), connecting them to text discussing parameter increases to 175B, 600B, and 1.6T. Other unhighlighted models listed in the table include BERT, DistilBERT, ALBERT, XLNet, RoBERTa, and T5-11B.}
  \label{fig:narrative-example}
\end{figure*}

Despite this critical role, data tables rarely stand alone; their full meaning is often co-constructed with the surrounding text through the narrative interplay between prose and tabular information (see Fig.~\ref{fig:narrative-example}).
This narrative interplay creates a major challenge: readers must continuously move between dense paragraphs and complex tables while manually aligning the narrative with the referenced data.
Such repeated back-and-forth navigation strains working memory and increases intrinsic cognitive load~\cite{sweller1988cognitive}.
To reduce the cognitive effort involved in relating narrative claims to the evidence that supports them, prior work has explored interactive reading interfaces that help readers make these connections more explicit.
For instance, systems have been developed to contextualize mathematical formulas~\cite{head2022math}, videos~\cite{kim2023papeos}, citation sources~\cite{chang2023citesee}, and terms~\cite{head2021augmenting}.
Regarding text and tables, existing efforts have proposed solutions to link sentences with table elements in web documents~\cite{badam2018elastic} and general PDFs~\cite{kim2018facilitating}, demonstrating that aligning textual references with tabular content can significantly improve both accuracy and efficiency within the sensemaking process.


However, creating meaningful connections between text and tables in scientific papers requires capabilities that go far beyond what existing methods currently support.
First, scientific writing relies on multiple types of links between text and tables, including semantic references, numerical comparisons, and structural cues, as illustrated in Fig.~\ref{fig:narrative-example}.
Existing approaches typically depend on syntactic features or surface-level semantic similarity to match sentences with individual table cells~\cite{badam2018elastic, kim2018facilitating}. 
Although these methods capture such explicit matches, they struggle to account for the rich and often implicit referencing patterns that characterize scientific narratives.
They are particularly limited when expressions require inference, such as \textit{``state-of-the-art''}, \textit{``an improvement of 3.7\%''}, or \textit{``the first two columns''}.

Second, to fully comprehend tabular data within a structured scientific narrative, readers must synthesize information across varying levels of text.
At the paragraph level, they need to understand how the narrative situates the role of a table within the broader argument. 
At the sentence level, they examine specific claims that point to particular results. 
When verifying a specific value or interpreting a detailed description, they must further inspect fine-grained textual spans and determine to which regions of the table these spans correspond.
This underscores the necessity for a text-to-table linking mechanism that supports progressive disclosure~\cite{shneiderman2003eyes} and integrates naturally into the reading experience.

To address these challenges, our work begins with a mixed-methods study aimed to gain a deeper understanding of how tables and text interact within scientific narratives.
Our content analysis of 132 paragraph-table pairs characterized the narrative interplay, identifying key linking mechanisms and a reference system anchored by \textit{mentions}. 
As atomic units linking text to tables, these mentions lay the foundation for our multi-granular alignment architecture. 
Complementing this analysis, semi-structured interviews with 12 researchers revealed the layered hierarchy of reading intents, ranging from information retrieval to narrative sensemaking.
These insights clarified how readers shift between different granularities during interpretation and directly informed the design requirements for our augmented reading interface.

We present \tool, an augmented reading interface designed to revive the narrative interplay between tables and text in scientific papers. 
Leveraging the general reasoning capabilities of large language model agents, we develop a multi-granular text-to-table alignment framework that automatically constructs a document-level linking schema. 
This schema is then rendered within an in-situ PDF interface, where progressive cascade activation reveals visual connections in a manner that supports smooth exploration from overview to detail.
We conducted a within-subject study with 24 participants to assess the overall performance of \tool.
Results show that \tool reduced perceived cognitive workload and improved reading efficiency, while also encouraging researchers to engage more deeply with results sections they might otherwise skip.
Interview feedback revealed an important tension: although the system accelerates evidence lookup, semantic understanding still requires active engagement from the reader.
Furthermore, our technical evaluation highlights that LLM agents can efficiently generate multi-granular links, yet occasional inaccuracies may still arise. 
A verification layer incorporating human-in-the-loop validation or automated consistency-checking may further mitigate these errors and ensure the stability of the linking schema in real-world scenarios.
In summary, our key contributions are as follows:
\begin{enumerate}
    \item Our formative study revealed the linking mechanisms, multi-granular alignment patterns, mention typology, and layered reading intents that characterize the narrative interplay between text and data tables.
    \item We present \tool, an augmented reading interface that uses an LLM-agent pipeline to construct a document-level linking schema and render it through cascade visual activation within scientific papers.
    \item We evaluated \tool through a user study, demonstrating its effectiveness and revealing meaningful user feedback for future refinement.
\end{enumerate}
\section{RELATED WORK}

We review prior work on interactive scientific reading systems, augmented reading interfaces, and methods for text-table alignment.
Bridging these foundations, our work introduces a reading support mechanism that uncovers the underexplored narrative interplay between text and data tables, making it explicit and accessible.

\subsection{Interactive Scientific Reading}
\label{ssec:interactive_scientific_reading}

Reading scientific papers is a cognitively demanding task~\cite{mayer2005cognitive,o1997comparison}.
Prior research has addressed this at the cross-paper level by organizing and synthesizing literature through citation contexts~\cite{Rachatasumrit2022citeread,chang2023citesee}, research threads~\cite{Kang2022Threddy}, and narrative building~\cite{zhang2025paperbridge,kang2023synergi,lee2024paperweaver}. 
In contrast, we focus on the internal complexity of a single paper, where readers must navigate dense text, formulas, and figures.

To address this, researchers have proposed various interactive augmentation methods, broadly categorized into three approaches. 
One major strategy is \textit{linking}, which creates semantic connections to reduce the effort of cross-referencing. 
For example, ScholarPhi~\cite{head2021augmenting} provides in-situ definitions of terms and symbols, and Papeos~\cite{kim2023papeos} links textual descriptions with video segments. 
A second strategy is \textit{augmentation}, which enhances visual or structural presentation to highlight key content. 
For instance, Scim~\cite{raymond2023scim} provides faceted highlights of salient paper contents to facilitate rapid skimming, while Charagraph~\cite{damien2023charagraph} and Statslator~\cite{damien2023statslator} convert statistical passages into interactive visualizations. 
Lastly, \textit{generation} leverages NLP techniques or large language models to produce simplified explanations or summaries, thereby lowering comprehension barriers. 
Paper Plain~\cite{Tal2023PaperPlain} produces simplified summaries and definitions, Qlarify~\cite{fok2024qlarify} combines retrieval and generation to produce expandable, layered summaries, and Shin et al.~\cite{shin2024papercard} turn papers into concise design cards via LLMs and text-to-image models to support quick understanding and communication.

In summary, prior systems have made significant progress in supporting terminology comprehension, structural navigation, and rapid summarization, effectively easing textual-level burdens. 
However, scientific argumentation often hinges on the narrative interplay between text and tables, where textual claims are substantiated by tabular evidence. 
How to automatically uncover and explicitly present these internal connections within a single paper remains an open challenge.

\subsection{Augmented Reading Interfaces}
\label{ssec:augmented_reading_interfaces}

To alleviate readers' cognitive load and enhance comprehension, researchers have sought to transform static reading into an interactive, exploratory process. 
This paradigm shifts the view of a document from a static information container to a dynamic interface designed to \textit{``encourage truly active reading''}~\cite{VictorBretExplorable}. 
Existing efforts toward this goal generally fall into two categories.

\textit{Natively Interactive Documents.}  
This first category of research focuses on creating entirely new document formats and authoring tools that support interactivity from the ground up. 
The core vision is to make interactivity an inherent part of the document itself.
For example, some approaches allow authors to embed interactive 3D scientific figures directly within articles~\cite{Newe2016Enriching}.
Other toolkits like Idyll~\cite{conlen2018idyll} and Living Papers~\cite{heer2023living} provide frameworks for creating \textit{``living documents''} that integrate dynamic visualizations, executable code, and interactive data. 
These systems demonstrate the vast potential for hands-on exploration and deeper understanding.
However, they are limited by the need for authors to adopt entirely new publishing paradigms and tools, which makes them unsuitable for augmenting the vast corpus of existing static literature.

\textit{Post-hoc Augmentation of Static Documents.}  
The second category, more closely related to our work, seeks to endow existing static documents (e.g., PDFs) with interactive capabilities by overlaying an additional layer without altering the original content. 
This paradigm resonates with a longstanding vision in HCI, tracing back to the notion of hypertext~\cite{Jeff1987Hypertext}, which imagined documents enriched with associative links beyond their static form. 
Subsequent research has instantiated this vision in different ways. 
For instance, Augmented Math~\cite{chulpongsatorn2023augmented} and Augmented Physics~\cite{gunturu2024augmented} employ Augmented Reality technologies to make static mathematical formulas and physics diagrams interactive and explorable.
These works prove the feasibility of enhancing specific domain content without modifying the original text. 


\subsection{Computational Methods for Text-Table Alignment}
\label{ssec:table_text_alignment}

Existing research has extensively explored modeling the relationship between tables and text at the algorithmic level~\cite{badaro2023transformers,qin2022survey}.
A major line of research focuses on enabling machines to answer questions or verify claims against tabular data. 
Models such as TaBERT~\cite{yin2020tabert}, TURL~\cite{deng2022turl}, and RAT-SQL~\cite{wang2019rat} employ schema linking techniques to map natural language to specific table cells, enabling precise data access. 
Similarly, systems like TabFact~\cite{chen2019tabfact} compare textual assertions with tabular evidence to determine veracity. 
Collectively, these approaches establish essential technical foundations and highlight the increasing prominence of pre-trained language models (PLMs) and LLMs in table understanding tasks.

In Visualization and HCI communities, several systems have explored text-data alignment using charts as the primary medium. 
For example, prior work has investigated crowdsourced text–chart linking~\cite{Kong2014Extracting}.
Other systems include authoring tools that maintain text-visual consistency and data accuracy, such as CrossData~\cite{crossdata2022chen} and EmphasisChecker~\cite{EMPHASISCHECKER2024kim}, as well as conversational interfaces like VizTA~\cite{vizta2025wang} that map  textual references to chart elements.
Related efforts on automated fact-checking over visual data~\cite{fu2024towards} further highlight interest in grounding textual statements in structured visual evidence. 
More closely related to our work are systems designed to reduce the cognitive load of cross-referencing text and tables during reading. 
Elastic Documents~\cite{badam2018elastic} operates on e-documents and treats the document itself as an interactive, collapsible medium capable of embedding contextual visualizations. 
The system uses traditional text preprocessing and set-similarity methods to match keywords, generating contextual visual representations that help readers quickly grasp data distributions.
Similarly, Kim et al.~\cite{kim2018facilitating} focus on augmenting PDF documents directly. 
Their system employs dependency-based syntactic analysis combined with similarity matching to construct bidirectional links between text and tables, supporting close-reading tasks such as evidence finding.

However, applying these existing methods directly to scientific literature presents limitations.
Our formative study revealed that deep text-table linking fundamentally relies on the synergy of three mechanisms: semantic, numeric, and structural linking. 
Current keyword or heuristic approaches often struggle to distinguish subtle N-gram variations, such as model variants, which are critical for accurate scientific reading.
These observations motivate us to revisit the academic reading scenario more closely.
We introduce \tool, a system powered by LLM-based agents with general reasoning capabilities to support multi-granular text-table alignment and progressive-disclosure interaction.
\section{Formative Study}

Tables play a central role in scientific narratives, yet their narrative interplay with text remains underexplored, and little is known about how readers practically engage with it.
To establish this problem space, we combined a content analysis of paragraph--table pairs that reveal narrative practices with a semi-structured interview study that surfaces reader perspectives, thereby identifying key design considerations.

\begin{figure*}[t!]
  \centering
  \includegraphics[width=\linewidth]{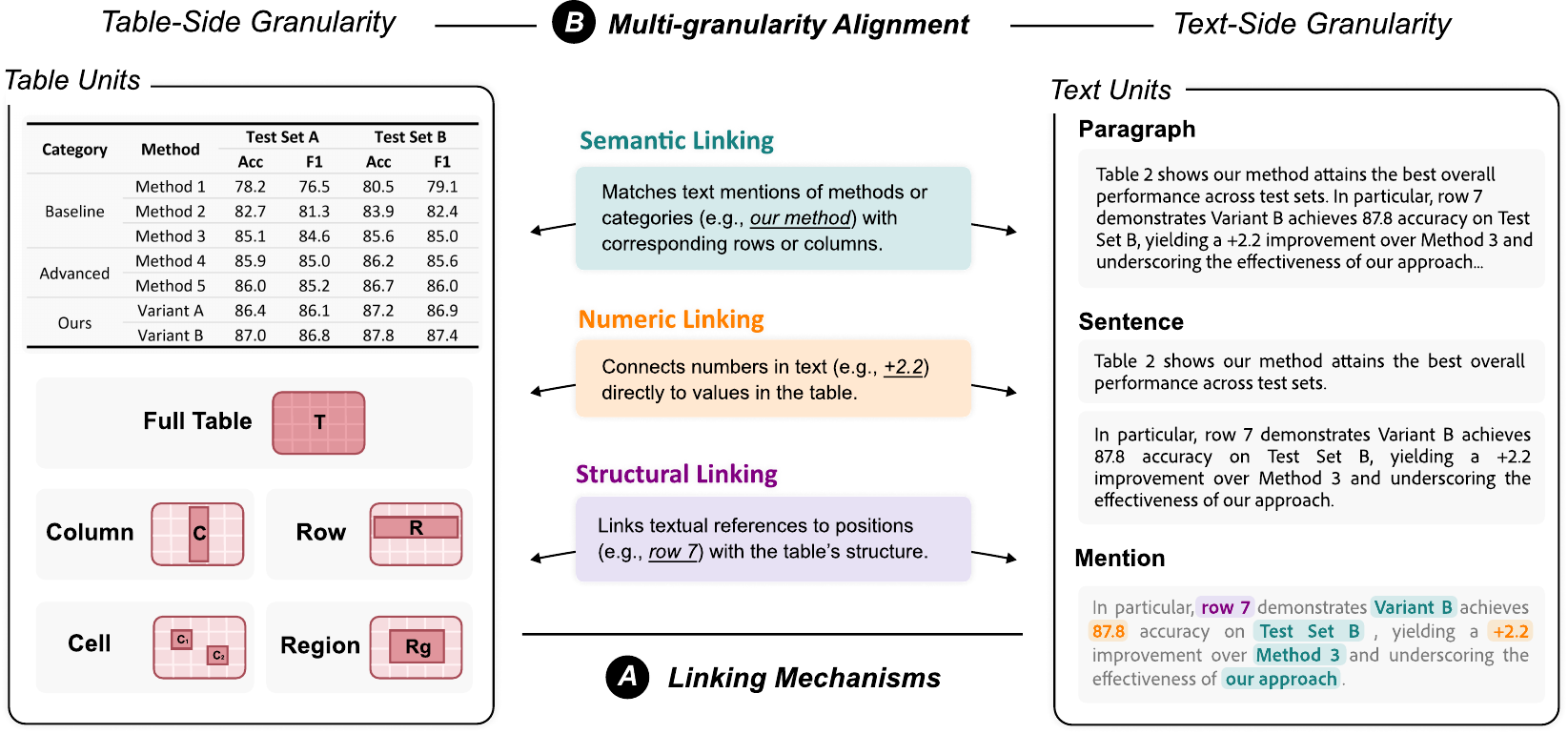}
  \caption{Alignment pattern between tables and text. 
Table-side granularity ranges from full tables to regions, rows, columns, and cells, while text-side granularity spans paragraphs, sentences, and mentions. 
The linking mechanisms include semantic, numeric, and structural connections that align units across the two modalities.}
\Description{A three-part diagram illustrating the Multi-granularity Alignment framework. 
The left panel, labeled Table-Side Granularity, presents a sample performance table listing categories, methods, and test set scores. Below this table are five icons representing table units: Full Table, Column, Row, Cell, and Region. Each icon uses a grid with red shading to visually define the specific unit (e.g., a vertical red strip for Column, a single red square for Cell). 
The right panel, labeled Text-Side Granularity, illustrates text processing levels. It shows a Paragraph of text describing the table, followed by a breakdown into Sentence level, and finally Mention level. In the Mention view, specific words are highlighted in different colors to represent extracted entities. 
The central panel, labeled Multi-granularity Alignment, connects the left and right panels via three distinct mechanisms. Top: Semantic Linking (in a teal box) connects text mentions like "our method" to table rows or columns. Middle: Numeric Linking (in an orange box) connects numbers in the text like "+2.2" to specific table values. Bottom: Structural Linking (in a purple box) maps positional references like "row 7" to the table structure. Arrows extend from these central boxes to the corresponding visual examples on the left and right sides.}
  \label{fig:alignment_pattern}
\end{figure*}

\subsection{Understanding Tables-Text Narrative Interplay in Scientific Papers}
\label{ssec:formative_survey}

Through a content analysis of paragraph-table pairs, we systematically organized our findings into three dimensions: the mechanisms of linking, the levels of granularity, and the typology of mentions.

\subsubsection{Content Analysis}

\revised{We focused our analysis on computer science, a field that frequently relies on tables to report configurations and experimental results.} 
We conducted a content analysis of 59 high-quality papers to characterize how tables and text are interwoven. 
These papers were drawn from best-paper selections at top-tier conferences and journals~\cite{aibestpapers2025,huangBestPaperAwards2025}, chosen for their exemplary writing and broad recognition within the research community.
\revised{To construct our corpus, we developed a script to automatically parse each paper, extract tables and full text, and retrieve candidate paragraph-table pairs using keyword-based heuristics.}
These candidates were then manually verified through a lightweight screening interface, yielding 132 validated paragraph–table pairs for subsequent analysis.
\revised{For the qualitative analysis, the first author performed inductive open coding on a subset of samples to surface initial categories of connections between text and tables. 
Particular attention was given to identifying key mention spans in the text, which are text units that refer to table content.
These preliminary categories were iteratively refined into a deductive codebook by two authors, who collaboratively identified mentions and aligned them with corresponding table elements.}

\subsubsection{Findings}
The analysis surfaced three complementary dimensions of table-text interplay: 
(i) the reasoning mechanisms that establish connections, 
(ii) the levels of granularity at which alignments occur, and 
(iii) the typology of textual mentions that act as anchors for linking.

\point{Linking Mechanisms.}
Building on prior work distinguishing explicit and implicit references~\cite{kim2018facilitating}, our analysis identifies three primary mechanisms that associate textual references with specific table elements (Fig.~\ref{fig:alignment_pattern}A).
\noindent\semantic{Semantic Linking} relies on a linguistic understanding, such as lexical matching, contextual interpretation, or synonymy, to establish a connection. 
For instance, a reference to \textit{``our model''} or \textit{``the baseline''} directs the reader to a specific row or column through conceptual reasoning.
\noindent\numeric{Numeric Linking} is grounded in explicit values or derived quantities, enabling a direct match. 
This involves mathematical operations like value look-up, comparison, or arithmetic computation. 
An example is a sentence mentioning \textit{``85.1''} or \textit{``improves by 4.4\%''}, which leads the reader to the corresponding data point.
\noindent\structural{Structural Linking} utilizes positional or structural cues rather than content values. 
This requires the reader to use reasoning about the layout of the table, such as row index, column order, or header labels. 
For example, \textit{``the first row''} directly guides the reader to a specific positional target.

\point{Multi-granularity Alignment.}  
Table-text alignments do not occur at a single level but span multiple granularities.  
As shown in Fig.~\ref{fig:alignment_pattern} (B), text can be decomposed into paragraphs, sentences, and mentions, 
while tables range from full tables to regions, rows, columns, and cells.  
Alignment is therefore inherently multi-granular: a paragraph may summarize an entire table, a sentence may involve multiple table units, and a mention may precisely point to a specific row, column, region, or even multiple cells.  
Such layered connections are essential for scientific narratives but also introduce complexity and potential ambiguity, as different levels may overlap and interact.  
This highlights the need for systems to support flexible navigation across granularities, enabling readers to shift smoothly between levels.  
On this basis, we next examine which textual elements serve as anchors to initiate alignment.

\begin{table*}[t]
\centering
\caption{Codebook for mention annotation in scientific papers.
We classify mentions into three categories (Entity, Value,
Structural) with representative subtypes.
Example sentences illustrate how mentions appear in text,
alongside the reasoning type required for alignment.}
\begin{tabular}{@{}l p{2.4cm} p{7.6cm} p{1.8cm}@{}}
\toprule
\textbf{Mention Type} & \textbf{Subtype} & \textbf{Example Sentence with \m{Mention}} & \textbf{Linking} \\
  &   &   & \textbf{Mechanism} \\
\midrule

\multirow{3}{*}{Entity Mention}
& Named Entity & \m{Method A} achieves the highest accuracy on... & \semantic{Semantic} \\
& Referential Entity & ..., and \m{this setting} uses a larger batch size for training... & \semantic{Semantic} \\
& Inferred Entity & Compared with \m{the strongest baseline}, our model delivers improvements on every dataset examined. & \semantic{Semantic}, \numeric{Numeric} \\
\midrule

\multirow{2}{*}{Value Mention}
& Raw Value & The F1 score reaches \m{0.92} on the validation set. & \numeric{Numeric} \\
& Derived Value & The full system outperforms the best ablation by \m{4.2\%}. & \numeric{Numeric} \\
\midrule

\multirow{1}{*}{Structural Mention}
& - & The results are shown in \m{the first row}. & \structural{Structural} \\
\bottomrule
\label{tab:mention_topology}
\end{tabular}
\end{table*}

\point{Mention Typology.}  
\textit{Mention} refers to a span of text denoting or pointing to an entity, a concept widely studied in tasks such as named entity recognition and coreference resolution~\cite{lee2017end,parsing2009speech}. 
Building on this notion, we adapt the term to the context of scientific writing, where mentions serve as the finest-grained textual pointers to tabular data.  
To capture this role, we developed a fine-grained classification that links each mention type to its characteristic alignment mechanism and granularity.  
In our corpus of 132 validated paragraph-table pairs, we identified a total of 1388 mentions.  
We classified them into three categories, with proportions observed in our data, as summarized in Table~\ref{tab:mention_topology}.  

\begin{itemize}
    \item \textbf{Entity Mentions (68.7\%)} point to entities represented in the table. 
    We distinguish three subtypes:  
    (i) \textit{Named Entities} (51.4\%) directly use surface forms such as terms or method names in tables;  
    (ii) \textit{Referential Entities} (13.8\%) require resolving contextual references (e.g., \textit{our method}, \textit{this setting}) before alignment; and  
    (iii) \textit{Inferred Entities} (3.5\%), which involve additional semantic reasoning combined with tabular evidence (e.g., \textit{the strongest baseline}, \textit{state-of-the-art}).  
    Resolving entity mentions therefore ranges from straightforward string matching to context-dependent reference resolution and inference grounded in table data.  
    
    \item \textbf{Value Mentions (30.8\%)} directly reference numerical information in a table. 
    We distinguish (i) \textit{Raw Values} (17.4\%), which correspond to cell entries (e.g., \textit{0.92}, possibly after rounding or unit conversion), and (ii) \textit{Derived Values} (13.4\%), which are computed from multiple cells (e.g., \textit{4.2\%} as a relative improvement).  
    Resolving raw values typically requires one-to-one cell lookup, while resolving derived values requires one-to-many reasoning across rows or columns.  
    
    \item \textbf{Structural Mentions (0.4\%)} reference structural aspects of a table rather than its content.
    Expressions such as \textit{the second row} translate directly into layout-based reasoning or positional indexing.
\end{itemize}

\subsection{Semi-Structured Interviews}
\label{ssec:formative_interview}

Building on the content analysis, we conducted semi-structured interviews to investigate how readers engage with tables in scientific papers. 
From these interviews, we derived a framework of reading intents characterizing how readers interpret and navigate text-table narrative interplay.

\subsubsection{Method}

We employed a purposive sampling strategy to recruit 12 early-career researchers (P1--P12; 7 male, 5 female) from computer science and engineering backgrounds.
They were selected to represent heavy consumers of papers featuring dense data tables, allowing us to deeply explore their specific needs, difficulties, and workflows. 
Participants were asked to walk through recently read papers, identify challenging moments when interpreting tables, and reflect on how they navigated between textual claims and tabular evidence. 
To further probe expectations, we introduced a lightweight prototype displaying paragraph-table pairs from our corpus, supporting mention-level highlights that triggered corresponding table elements.
This early form of text-table interaction served as a stimulus to elicit feedback on design opportunities.  
\revised{All sessions were recorded with participants' consent and automatically transcribed. 
We then conducted an inductive thematic analysis: the first author performed open coding, which was iteratively refined through discussions with a second author to derive the final reading intents and design requirements.}

\subsubsection{Findings: A Framework of Reading Intents}

\begin{table*}[!t]
\centering
\caption{Mapping layered reading intents to design requirements.}
\begin{tabular}{m{0.22\linewidth} m{0.32\linewidth} m{0.38\linewidth}}
\toprule
\textbf{Reading Intent Layer} & \textbf{Example Behaviors or Needs} & \revised{\textbf{Design  Requirements}} \\
\midrule

\textbf{Low-level} \newline \textbf{Retrieval \& Alignment} 
& $\diamond$ Locate specific values  \newline
$\diamond$ Manage cross-page navigation \newline
$\diamond$ Match mentions across text and table 
& \revised{\textbf{DR1}: Reconstruct text-table narrative interplay via narrative-driven lookup to bridge the gap between text and structured data.}\\

\midrule
\textbf{Mid-level} \newline \textbf{Integration \& Verification} 
& $\diamond$ Extract entities and infer relations \newline
 $\diamond$ Verify textual claims with tabular data 
& \textbf{DR2}: Facilitate multi-granular alignment for efficient evidence verification and synthesis without losing granular details.\\

\midrule
\textbf{High-level} \newline \textbf{Narrative Sensemaking} 
& $\diamond$ Follow author’s argumentative flow \newline
$\diamond$ Distinguish main conclusions from details   
& \textbf{DR3}: Synchronize evidence with the narrative flow to preserve continuity of argumentation. \\

\midrule
\textbf{Across all levels} 
& $\diamond$ Maintain lightweight interaction \newline
$\diamond$ Avoid visual overload  
& \textbf{DR4}: Disclose cues on demand with minimal disruption, ensuring augmentation remains flexible. \\

\bottomrule
\end{tabular}
\label{tab:intent2design}
\end{table*}

From our formative interviews, we identified a layered structure of reading intents when participants engaged with tables in scientific papers. 
These intents span from low-level information retrieval, to mid-level integration, and to high-level narrative understanding.  

\point{Low-Level Intents: Retrieval and Alignment.}
A primary challenge for many participants was to \textit{locate specific values} or \textit{match entities} between text and tables.
For instance, P5 noted that \textit{``my biggest difficulty is flipping back and forth, I have to go back to check what a method means''}, highlighting the friction caused by cross-page navigation and unfamiliar terminology. 
P8 described a similar struggle when encountering percentages, explaining that \textit{``I naturally want to verify the value difference, but I cannot directly find it and have to calculate by myself.''}
Several participants also reported forgetting information quickly or taking screenshots to keep track, suggesting that even low-level retrieval often comes with high operational cost.

\point{Mid-Level Intents: Integration and Verification.}
After locating relevant information, participants attempted to \textit{verify claims} in the text against the table or to \textit{extract takeaways} such as which method performs best. However, claim verification was often described as an ideal but rarely completed practice. 
As P4 emphasized, \textit{``I would not really verify every single data fact''}. 
P9 similarly complained that \textit{``I need to match each method mentioned in the paragraph to the rows in the table one by one, and that is really laborious''}.
Still, some participants stressed the importance of reasoning about differences or improvements. 
For example, P7 explained, \textit{``I care more about why the performance is better, the reasons behind it, rather than the exact number''}.
This indicates that mid-level intents involve both the need and the burden of validation and synthesis.

\point{High-Level Intents: Narrative Sensemaking.}
At a higher level, participants expressed the need to \textit{follow the author’s narrative flow} and build a coherent understanding of the study. 
P4 wished that highlighting could surface the most critical conclusions rather than peripheral details, remarking that \textit{``I hope what is highlighted is the article’s main conclusions, not all the trivial mentions''}. 
P7 further pointed out that he paid more attention to failure cases, stating that \textit{``I rarely care about successful cases, because they may be exaggerated; the reasons for failures are often more insightful''}.
Others requested richer contextual anchoring.
P8, for example, suggested that \textit{``it would be very useful if the system could link models directly back to their first occurrence or to the reference''}.
Overall, high-level intents reflect the need to synthesize scattered mentions into a structured whole while preserving the continuity of argumentation.

In summary, low-level intents center on locating numbers and entities, mid-level intents focus on verifying claims and extracting key takeaways, and high-level intents emphasize narrative and contextual sensemaking. 
Yet not all intents are equally pursued in practice: while meticulous claim verification is often skipped, participants consistently valued efficient lookup, clarity of takeaways, and continuity of narrative. 
This layered view highlights diverse but recurring challenges, providing a structured basis for informing our design rationales in the next section.

\subsection{Design Requirements}

{Grounded in our content analysis and the reading intents identified in the interviews, we derived four design requirements for augmenting table-text narrative interplay in reading scientific papers. 
Table~\ref{tab:intent2design} maps the identified reading intents and associated reader behaviors to our core design requirements, illustrating the transition from formative insights to system support.

\begin{itemize}
    \item \textbf{DR1: Reconstruct text-table interaction via narrative-driven lookup.}
    To bridge the cognitive gap between text and structured tables, the system should prioritize aligning table units from the text side to support narrative reading. 
    This design follows the reader's natural habit of using text as the primary entry point, allowing them to instantly locate tabular facts without breaking the narrative flow.
    
    \item \textbf{DR2: Support layered alignment across granularities.} \revised{Informed by the multi-granular nature of text-table connections, the system should replace distracting single-layer highlighting with flexible, layered support. 
    This enables readers to seamlessly shift between granularities to verify specific values at the micro-level while maintaining a macro-level overview for key takeaways.}

    \item \textbf{DR3: Present in-situ evidence in sync with the narrative progression.}
\revised{To preserve narrative coherence, the system should display table evidence only when it becomes relevant in the text. 
The system should provide situated visual augmentation on text and table to maintain the continuity of the author’s argument.}

    \item \textbf{DR4: Disclose cues on demand with minimal disruption.}  
Across all reading layers, augmentation should remain lightweight and avoid visual overload. Strong or persistent highlights distracted participants from the main text. On-demand cues (e.g., triggered by hover or click) allow flexible interaction and support different levels of engagement without imposing a rigid workflow.
\end{itemize}

\section{TableTale}
\label{sec:system}

\tool is an augmented reading tool designed to help readers understand and verify the text-table narrative interplay in scientific papers. 
It integrates 
(i) an automatic offline pipeline that constructs a document-level \textit{linking schema} capturing multi-granular text-table alignment, and 
(ii) an in-situ PDF interface that operationalizes the schema as \textit{table augmentation} through \textit{progressive cascade activation}, guiding readers from high-level paragraph context down to fine-grained table units during the reading process.

\begin{figure*}[t!]
  \centering
  \includegraphics[width=0.99\linewidth]{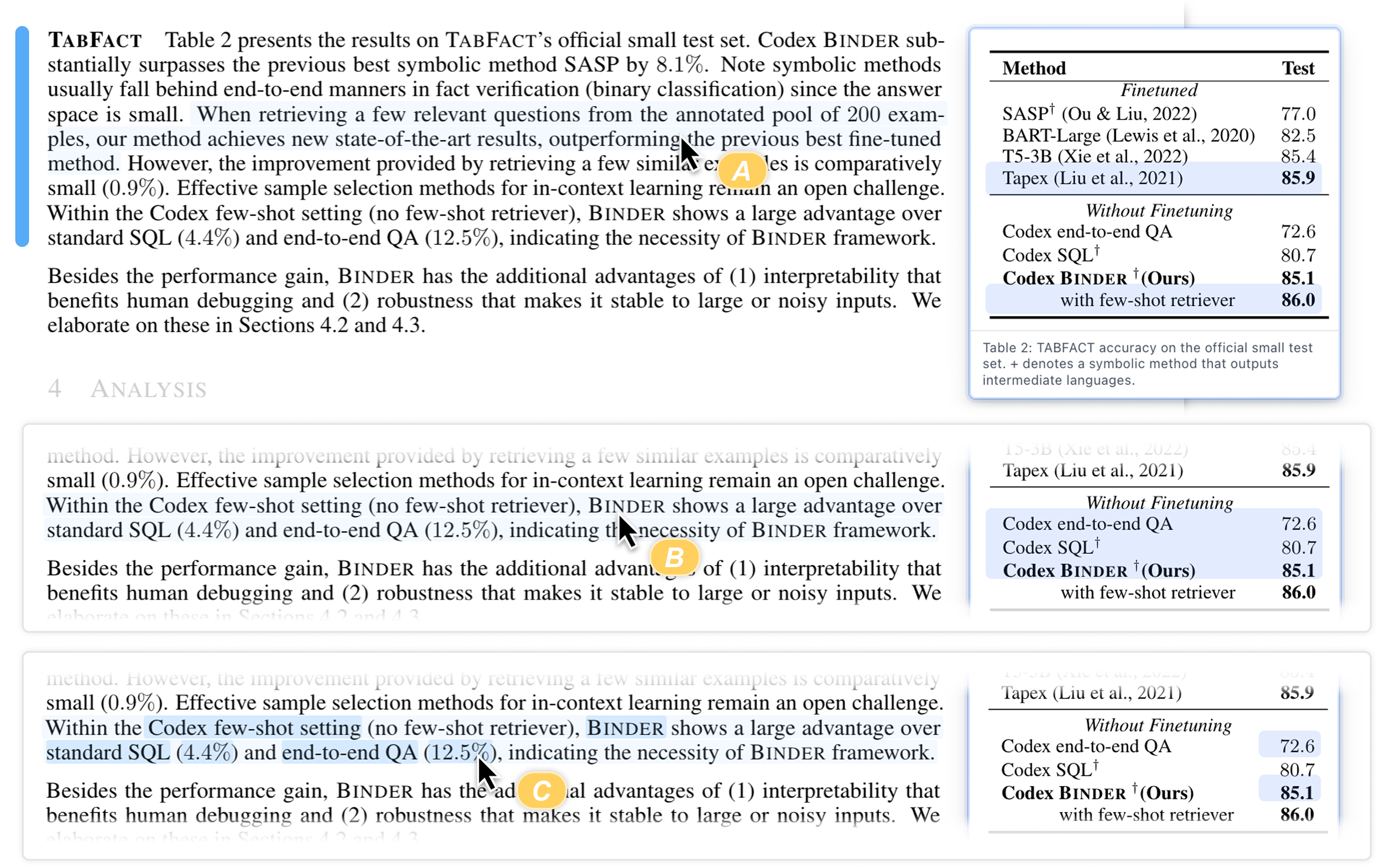}
  \caption{\tool interface with Progressive Cascade Activation.
  (a) A sidebar cue indicates paragraphs containing table references; activating it anchors the corresponding table next to the paragraph even if the table is off-screen.
  (b) Hovering over a sentence highlights only the relevant cells, rows, columns, or regions in the anchored table.
  (c) Clicking a sentence reveals its mentions, and hovering over a mention (e.g., ``12.5\%'') further highlights the exact evidence cells (e.g., 85.1 and 72.6).}
\Description{A three-part composite image labeled A, B, and C showing interaction stages within a document interface. The general layout displays a research paper text column on the left and a floating data table titled Table 2 on the right. 
In stage A, a vertical blue bar appears in the left margin next to a paragraph starting with TABFACT, visually connecting the text to the floating table. 
In stage B, a yellow cursor labeled B hovers over the sentence starting 'Within the Codex few-shot setting.' The entire sentence is highlighted with a light blue background. Simultaneously, the bottom three rows of the table (labeled Codex end-to-end QA, Codex SQL, and Codex BINDER) are highlighted in matching light blue to show the row-level correspondence. 
In stage C, a yellow cursor labeled C hovers specifically over the text '12.5 percent' within the same sentence. This text span is highlighted in a darker blue. Correspondingly, two specific cells in the table containing the values 72.6 and 85.1 are highlighted in blue, indicating the exact source data used to calculate the percentage in the text.}
  \label{fig:interface}
\end{figure*}

\subsection{User Interface}
\label{ssec:user_interface}

The interface of \tool builds on two core mechanisms: 
(1) \textit{progressive cascade activation}, which incrementally triggers these highlights from paragraph- to sentence- to mention-level text references, and
(2) \textit{table augmentation}, which highlights relevant table regions based on the linking schema.

\subsubsection{Progressive Cascade Activation}

Progressive cascade activation enables table highlights to be triggered in sync with the narrative flow, allowing readers to progressively uncover evidence as they move from one paragraph or sentence to the next. 
It also supports a coarse-to-fine drill-down pathway, guiding readers from broad contextual cues to increasingly precise visual references within the table. 
The interaction consists of three levels, and are all rendered in-situ on the PDF text layer and driven by the offline-generated \textit{linking schema}. 
Fig.~\ref{fig:interface} illustrates this mechanism.

\begin{itemize}
  \item \textit{Paragraph-Level:} When a paragraph contains a table reference, a sidebar cue appears alongside it. 
  Clicking the bar activates the linking augmentation, enabling interactive highlights for subsequent text levels. 
  If the referenced table is currently off-screen, \tool additionally anchors a mirror copy next to the paragraph (Fig. 4a). This ensures readers can access evidence without flipping pages or manual searching.

  \item \textit{Sentence-Level:} Once the paragraph level is activated, hovering over a sentence highlights the sentence itself and, simultaneously, the anchored table as a dynamic canvas, where only the units (cells, rows, columns, or regions) relevant to that sentence are emphasized (Fig.~\ref{fig:interface}b). 
The highlights update as the reader moves between sentences, reducing visual clutter and maintaining focus.

  \item \textit{Mention-Level:} When the reader clicks on an activated sentence, its internal mentions are revealed. 
    Hovering over a mention (e.g., ``12.5\%'') then highlights the exact scope for the mention on the table (e.g., the cells for 85.1 and 72.6) (Fig.~\ref{fig:interface}c). 
    Through this drill-down, progressive cascade activation preserves high-level readability while enabling precise verification at finer levels.

\end{itemize}

\subsubsection{Table Augmentation}

Table augmentation in \tool focuses on two primary aspects: spatial accessibility and contextual focus. 
The system manages the placement of tables to ensure they remain within the reader’s viewport, while simultaneously highlighting internal units to direct attention to the specific regions relevant to the current narrative.

\textbf{Adaptive Table Placement.} \tool supports two rendering modes~\cite{kim2018facilitating} depending on whether the referenced table is visible. 

\begin{itemize}
    \item \textit{Anchored Rendering}. 
    When a table is off-screen, a mirror copy is anchored next to the active paragraph, allowing readers to access evidence without leaving the textual context. 
    Fig.~\ref{fig:interface} (A-a) illustrates the rendering of an anchored mirror table, which appears adjacent to the page upon hovering over the specific sentence reference.
  \item \textit{In-situ Rendering}. 
  When the table is  visible, highlights are directly applied to the original table, preserving spatial continuity and avoiding redundancy. 
  Fig.~\ref{fig:teaser} (b-d) illustrates examples of highlight forms within in-situ tables.
\end{itemize}

\textbf{Highlight Rendering.} 
When triggered from the text side, highlights appear on the table as subtle bounding boxes overlaid directly on the table elements. 
The offline-generated \textit{linking schema} specifies both their granularity (row, column, region, or cell) and location, ensuring precise correspondence between text and tabular evidence. 
Fig.~\ref{fig:teaser} illustrates examples of region-level (a), column-level (b), row-level (c), and cell-level (d) highlights, respectively.

\begin{figure*}[t!]
  \centering
  \includegraphics[width=0.99\linewidth]{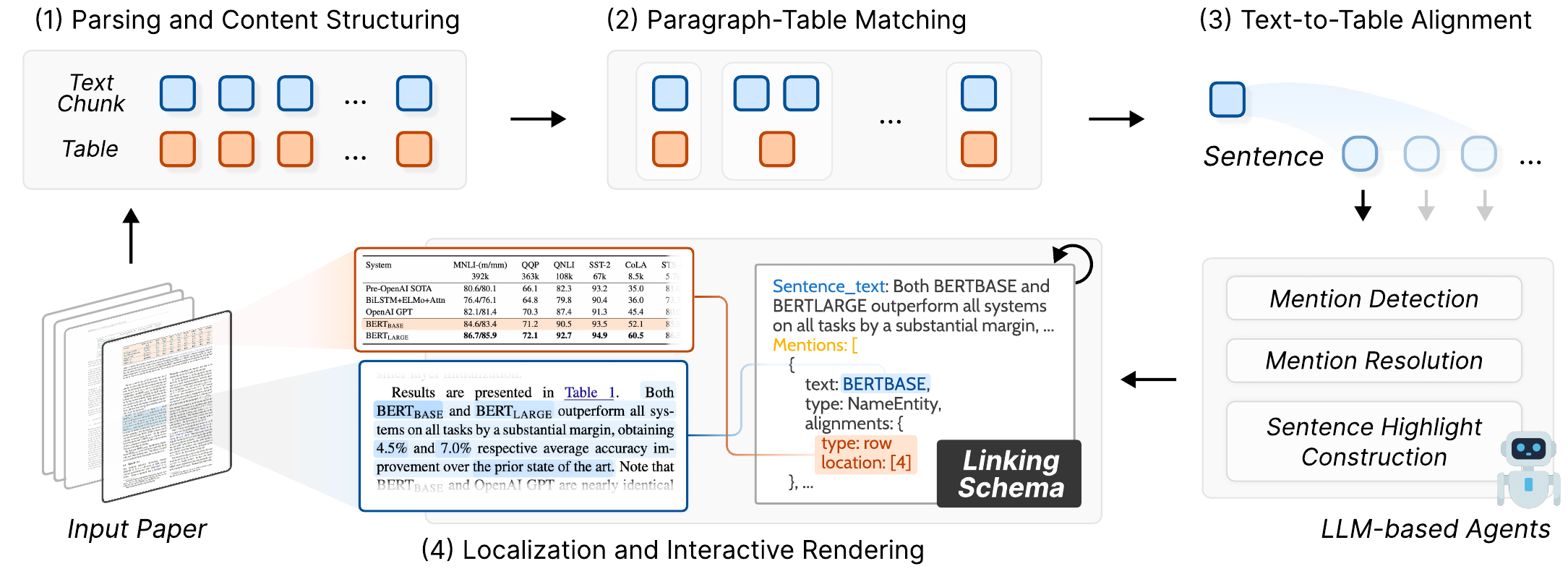}
  \caption{%
Pipeline of \tool. The system processes an input paper through four stages: (1) parsing and structuring, (2) paragraph-table matching, (3) fine-grained text-to-table alignment via multi-agent collaboration, and (4) localization and interactive rendering.%
}
\Description{A four-stage workflow diagram flowing from left to right, starting with an icon of input papers and ending with an interactive user interface.
The first stage, Parsing and Content Structuring, shows the input paper decomposing into abstract representations, where text chunks are visually encoded as blue squares and tables as orange squares.
The second stage, Paragraph-Table Matching, depicts these abstract squares being grouped together to show associations between specific text chunks and tables.
The third stage, Text-to-Table Alignment, shows a single blue text chunk expanding into a sequence of circles representing sentences. These feed into a Multi-Agent block, illustrated with a robot icon, which lists three internal modules: Mention Detection, Mention Resolution, and Sentence Highlight Scope Construction.
The fourth stage, Localization and Interactive Rendering, presents a detailed close-up of the final interface. On the left is a results table containing model performance data. On the right is a text paragraph analyzing the table. A highlight on the word BERT-BASE in the text is connected by a curved line to the corresponding row in the table. An overlay box labeled Linking Schema displays the underlying JSON structure for this connection, showing fields for text, entity type, and row location alignments.}

  \label{fig:pipeline}
\end{figure*}

\subsection{System Architecture}
\label{ssec:system_architecture}

\tool features a position-aware, multi-granular text-to-table alignment framework that establishes a closed-loop mechanism to augment static PDFs with interactive linking capabilities. 
At the core of this architecture is a document-level linking schema, a foundational hierarchical data model designed to encapsulate alignment results and drive multi-granular rendering within the interface.
As illustrated in Fig.~\ref{fig:agent}A, the schema is organized hierarchically: a \textit{document} comprises multiple \textit{paragraph-table pairs}, which are further decomposed into constitutive \textit{sentences}. 
At the finest level of granularity, each sentence encodes specific \textit{mentions} and their associated \textit{alignments} to table elements. 
The following subsections delineate the computational pipeline developed to instantiate this schema through four distinct stages.

\subsubsection{Document Parsing and Content Structuring}  
The pipeline begins by decomposing the source PDF into structured, addressable units. 
To ensure high-fidelity, position-aware operations, we utilize the TextIn~\cite{textin} to extract data tables and text blocks with preserved spatial metadata. 
Specifically, we segment the textual content into paragraphs  
($P = \{P_{1}, P_{2}, \dots, P_{n}\}$) and extract data tables  
($T = \{T_{1}, T_{2}, \dots, T_{m}\}$), preserving both their semantic structures and spatial metadata.  
For tables, this representation goes beyond cell values and layouts (e.g., merged cells) to include captions, titles, and page-level coordinates, together with cell-level bounding boxes obtained through table structure recognition.  
All extracted tables are stored in HTML format, with every cell assigned a unique identifier (e.g., \texttt{r1c3} for the cell with row index 1 and column index 3) encoding its structural position.

\subsubsection{Paragraph-Table Matching}
The next stage identifies candidate paragraphs that discuss specific tables to focus the alignment process. 
Since only a subset of the document typically references tabular data, processing the entire text would introduce unnecessary computational noise.
We identify paragraphs containing explicit table references (e.g., ``Table~\#'', ``Tab.~\#'') using regular-expression patterns to extract seed text chunks. 
To handle structural artifacts such as pagination or layout breaks, we employ a heuristic merging strategy with neighboring chunks to reconstruct the full paragraph context. 
This process yields validated paragraph-table pairs $(P_i, T_j)$ indicating that paragraph $P_i$ refers to table $T_j$. 
These pairs provide the input for subsequent sentence- and mention-level alignment.

\begin{figure*}[t!]
  \centering
  \includegraphics[width=0.99\linewidth]{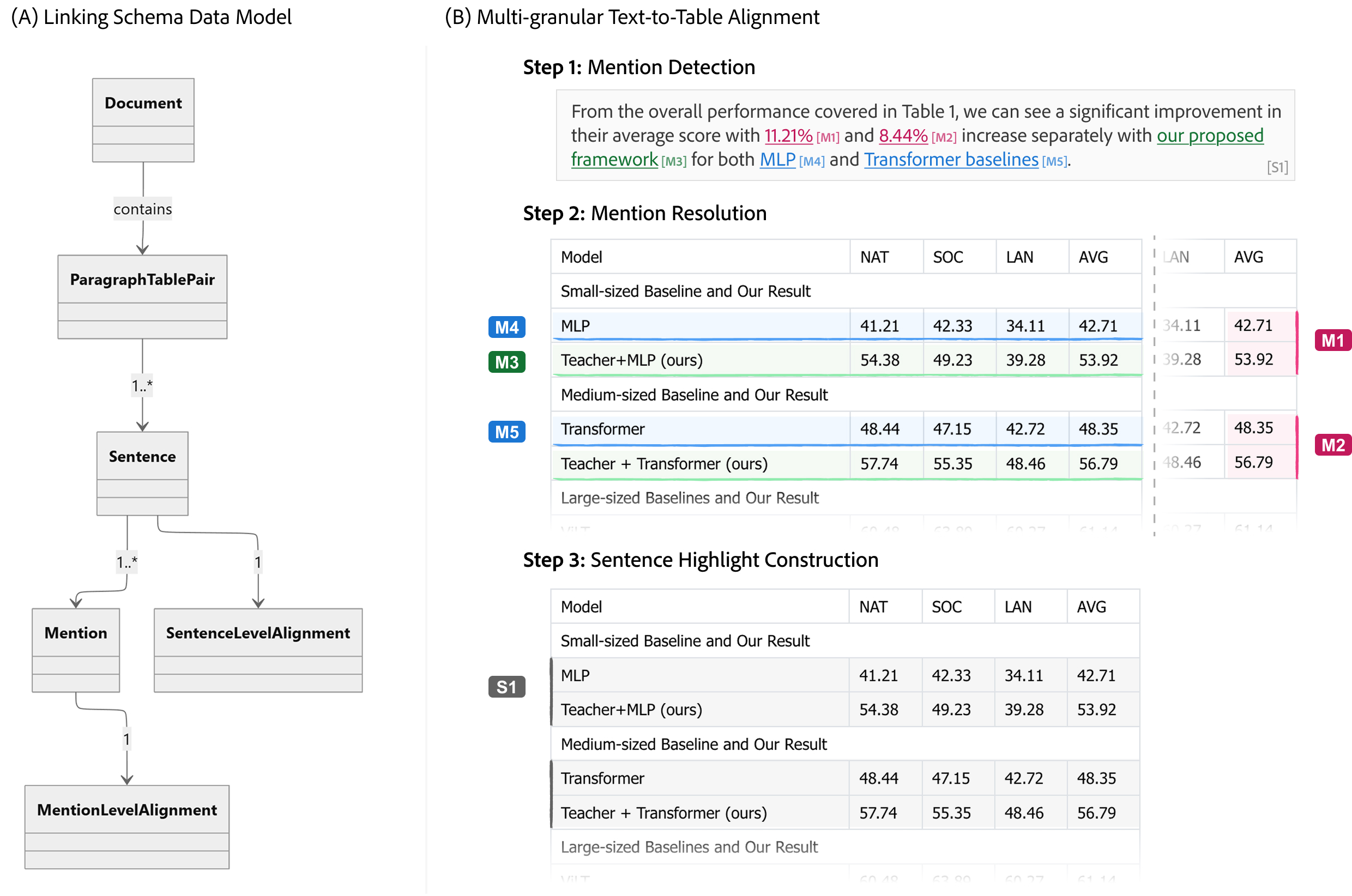}
  \caption{Overview of our multi-granular text–table alignment framework. 
(A) Linking schema data model that organizes paragraphs, sentences, mentions, and their alignment targets. 
(B) Bottom-up alignment procedure illustrated with an example sentence, showing how fine-grained mention links are progressively merged into sentence-level table regions.}
  \Description{A two-part figure illustrating the system's data structure and workflow. Panel A shows a vertical class diagram representing the Linking Schema Data Model. At the top, a Document box connects downward via a "contains" relationship to a ParagraphTablePair box. This pair connects via a one-to-many relationship to a Sentence box. The Sentence box branches into two lower boxes: a Mention box (via a one-to-many relationship) and a SentenceLevelAlignment box (via a one-to-one relationship). Finally, the Mention box connects downward to a MentionLevelAlignment box via a one-to-one relationship. Panel B illustrates the Multi-granular Text-to-Table Alignment in three steps using a specific example. Step 1, Mention Detection, shows a text sentence about performance improvements with specific percentages (11.21 percent, 8.44 percent) and model names (our proposed framework, MLP, Transformer baselines) highlighted in color. Step 2, Mention Resolution, displays a table with performance metrics (NAT, SOC, LAN, AVG) for Small-sized and Medium-sized baselines. Labels M3, M4, and M5 identify specific rows for Teacher+MLP, MLP, and Transformer, respectively. Brackets M1 and M2 on the right side of the table group rows corresponding to the percentage increases mentioned in the text. Step 3, Sentence Highlight Construction, shows the final alignment where the relevant table section (the Small-sized Baseline rows) is highlighted and labeled S1.}
  \label{fig:agent}
\end{figure*}

\subsubsection{Multi-Granular Text-to-Table Alignment}

\tool performs text-table alignment through a multi-granular, bottom-up pipeline that connects natural language expressions to the appropriate table elements. 
The process begins by segmenting each paragraph $P_i$ into a set of constitutive sentences $S_i = \{s_1, s_2, \dots, s_u\}$. 
Each sentence $s_k$ is then paired with its associated table $T_j$ to form sentence-table pairs $(s_k, T_j)$, which serve as the fundamental units for alignment.
As illustrated in Fig.~\ref{fig:agent}, the framework integrates a document-level linking schema with a three-step approach, where three specialized agents operate in a progressively integrative manner.

\paragraph{Mention Detection Agent.}
Given a sentence-table pair \((s, T)\), this agent identifies all natural language expressions in the sentence $s$ that may refer to the content of table $T$. 
It produces a structured set of mention candidates,
\(\mathcal{M} = \{\, (\text{text},\, \text{span},\, \text{type}) \,\}\),
where each mention consists of a surface text form, 
a character-level span in the sentence, 
and a semantic type as defined in Table~\ref{tab:mention_topology}.
For example, in Fig.~\ref{fig:agent}B (Step~1), the agent identifies five mentions from sentence $S_1$, including \textit{Derived Value} (\(M_1, M_2\)), \textit{Referential Entity} (\(M_3\)), and \textit{Named Entity} (\(M_4, M_5\)) categories.

\paragraph{Mention Resolution Agent.}
This agent grounds each detected mention to the most semantically relevant element in the table. 
Given a mention set \(\mathcal{M}\) and table \(T\), the agent assigns each mention \(m \in \mathcal{M}\) an alignment target drawn from cells, rows, columns, or table regions, expressed as \(\mathcal{A}(m) \in \{\text{cell}, \text{row}, \text{column}, \text{region}\}\). 
The resolution procedure combines lexical matching, numerical reasoning, and structural constraints from table \(T\) to determine a valid grounding.
As shown in Fig.~\ref{fig:agent}B (Step~2), the agent resolves mentions \(M_1\) and \(M_2\) by computing the corresponding numeric differences (e.g., mapping ``11.21\%'' to the two source cells ``42.71'' and ``53.92'').
Mentions that cannot be grounded to any element are discarded at this stage.

\paragraph{Sentence Highlight Agent.}
To formalize the mapping between the entire sentence $s$ and its collective evidence in table $T$, this agent merges fine-grained mention targets into higher-level regions \(\mathcal{R}(s)\).
By prioritizing structural integrity over exhaustive detail, the agent aggregates overlapping or adjacent targets into cohesive structures such as rows, columns, or regions.
As illustrated in Fig.~\ref{fig:agent}B (Step 3), the agent consolidates five individual mention alignments into two highlight regions for sentence \(S_1\).

Together, these three agents form a coordinated pipeline for deriving multi-granular text-table alignments. 
Their performance relies on the in-context learning capability of large language models, guided by a small set of few-shot examples. 
All agents are implemented using OpenAI’s \texttt{gpt-4o} model as the backbone for reasoning and alignment decisions. 
This process ultimately yields a document-level linking schema that supports downstream rendering in the reading interface.

\subsubsection{Localization and Interactive Rendering}  
The final stage of the pipeline converts the computed alignments into a responsive user experience. 
Upon opening a PDF document, the interface is initialized, driven by the pre-computed document-level linking schema.
The schema serves as the foundational data layer, triggering interactive elements on the text side and rendering visual highlights on the table side.
These interactive components are seamlessly synchronized via the \textit{progressive cascade activation} mechanism.
By leveraging the spatial metadata captured during parsing, \tool maps the bounding boxes of all addressable units onto a normalized coordinate system based on percentage-based offsets relative to page dimensions.
This rendering strategy guarantees that overlay highlights remain consistently and accurately aligned with the original PDF content, regardless of viewport scaling or adjustments.

\subsection{Implementation Details}

\tool was implemented as a standalone web application. 
The front-end interface was built with JavaScript, CSS, and the Vue framework, and PDF rendering was powered by PDF.js. 
Backend services and LLM-based functions were implemented in Python using Flask. 
For basic NLP tasks, we employed the \texttt{spaCy} \texttt{en\_core\_web\_sm} model for sentence segmentation 
and BeautifulSoup4 for parsing HTML tables. 
To improve responsiveness, the system dynamically caches parsing results and stores the computed linking schema, 
avoiding redundant calls during repeated access to the same paper.

\section{Technical Evaluation}

We evaluated the two critical components of our multi-granular text-table alignment pipeline: the \textit{mention detection agent} and the \textit{mention resolution agent}. 
In the bottom-up alignment process, these components directly dictate the overall accuracy of the generated links. 
To ensure consistency, all experiments were conducted using the same LLM backend configuration as the deployed system.

\subsection{Dataset and Setup}

We conducted our evaluation using a subset of paragraph-table pairs drawn from the corpus developed in our formative study. 
Each pair in this corpus includes manually verified mention annotations and alignment labels, serving as a reliable gold standard for assessing system performance. 
Following prior work highlighting the variability of table sizes in scientific papers~\cite{zhong2020PubTabNet}, we categorized tables by their area (rows $\times$ columns) into three groups based on corpus-level quantiles:
(1) \textit{simple} tables with an area of up to 48 cells,  
(2) \textit{standard} tables with an area between 48 and 90 cells, and  
(3) \textit{complex} tables exceeding 90 cells.  
To account for the structural diversity of tables, we drew 25 samples across different levels of table complexity as the evaluation set.

\subsection{Evaluation Metrics}

For mention detection, we followed evaluation practices commonly used in named entity recognition, where a predicted span was counted as correct if its character-level Intersection-over-Union (IoU) with the gold span satisfied \(\text{IoU}\ge0.5\). 
Precision, Recall, and F1 were computed using standard definitions.
For mention resolution, we evaluated whether each detected mention was correctly aligned to its corresponding table element.
We adopted a strict all-or-nothing metric that considered an alignment correct only when the predicted region covered exactly the same table area as the gold annotation in spatial extent.

\subsection{Results}

The mention detection agent achieved a precision of 82.3\%, a recall of 86.3\%, and an F1 of 84.3\%. Meanwhile, the mention resolution agent reached an accuracy of 75.4\%.
Our findings validate the feasibility of employing LLM-based agents for fine-grained text-table alignment, while also revealing that occasional inconsistencies or hallucinations may still arise. 
In practice, these findings underscore the necessity of a verification layer, employing either human-in-the-loop validation or automated consistency checks, to stabilize the linking schema prior to rendering.
While mention detection performance remained stable across table complexities, mention resolution accuracy decreased substantially from simple (87.9\%) to standard (73.1\%) and complex tables (62.0\%).
This degradation reflects the increased difficulty of resolving ambiguous references and nested structures as tables become more complex.
\section{User Study}
We conducted a within-subject user study with 24 participants to evaluate the usability and effectiveness of \tool in supporting text-table reading within scientific papers. 
Specifically, our study addressed two key research questions:
\begin{itemize}
    \item \textbf{RQ1:} How effectively does \tool assist readers in understanding and integrating information from tables?
    \item \textbf{RQ2:} What are the potential benefits and drawbacks of using such a system for scientific paper reading?
\end{itemize}

\subsection{Study Design}

\subsubsection{Conditions}
All participants experienced two conditions.
(1) \textit{Basic}, an enhanced PDF.js reader that supports paragraph-level table linking, anchoring each referenced table next to its corresponding paragraph. 
This condition also retains standard PDF features such as text selection, zooming, and keyword search (CTRL-F).  
(2) \tool, the full-feature prototype as described in Section~\ref{sec:system}.

\subsubsection{Participants}
We recruited 24 early-stage researchers (13 male, 11 female; aged 22--28) through an open call at a local university, denoted as U1--U24.
Participants included Master’s students and junior Ph.D. students, primarily from computer science-related fields (e.g., data science, NLP, visualization). 
Aligning with prior work~\cite{damien2023charagraph, kim2023papeos}, we targeted this population because their regular engagement with scientific literature makes them ideal candidates to evaluate the utility of \tool.
To ensure familiarity with the domain conventions, the inclusion criteria required participants to have prior experience reading AI-related papers, specifically those employing data tables to present numerical evidence.
All participants were fluent in English and possessed high data literacy. 
Their self-reported paper reading frequency varied, with 8 participants reading daily, 13 weekly, and 3 monthly. 
Upon completion, each participant received a cash compensation equivalent to USD 10 for their time.
The study was conducted in a one-on-one, in-person setting, and the study protocol was approved by the Institutional Review Board of the university.

\subsubsection{Paper Excerpts}
We selected two authentic paper excerpts as experimental materials: 
(1) \textit{Binding Language Models in Symbolic Languages}~\cite{cheng2023binding} (ICLR 2023 Oral), and 
(2) \textit{Phrase-Based \& Neural Unsupervised Machine Translation}~\cite{lample2018phrase} (EMNLP 2018 Best Paper).
Each excerpt consisted of two consecutive and self-contained paragraphs from the results section, with each paragraph linked to a different table. 
The selected excerpts were comparable in length and had been pre-tested with three pilot participants to ensure comparable difficulty in terms of reading time and task completion. 
To ensure participants could quickly grasp the context of each paper, we also provided a brief reading primer that introduced the motivation and contributions in accessible language.

\subsubsection{Tasks}
For each paper excerpt, we designed four comprehension tasks, each based on a specific, table-related sentence within the excerpt. 
The tasks required participants to articulate their understanding by verbally validating a claim in the sentence using information from the corresponding table. 
This design went beyond simple correctness to capture the participants' reasoning processes.
Specifically, for each sentence, participants were asked to:

\begin{itemize}
    \item \textit{Explain Semantics Meaning}: Describe the meaning of the sentence and its relation to the table.
    \item \textit{Identify Objects}: Point out the specific table objects (e.g., rows, columns) that the sentence refers to.
    \item \textit{Verify Data Claim}: Verify and confirm the numerical values or data claims mentioned in the sentence.
\end{itemize}

\subsubsection{Ordering}
Text excerpts and conditions were presented with an order following a balanced Latin square design.
We ensured that (1) each participant saw each paper excerpt exactly once; (2) each condition was paired with exactly one paper excerpt; and (3) a paper excerpt appeared exactly once at every possible condition every four participants.
As such, half of the participants saw the same text excerpt in the \textit{basic} condition, while the other half saw it in the \tool condition. 
Additionally, we balanced the order of the conditions: half of the participants started with the \textit{Basic} condition while the other half started with the \tool condition.

\subsubsection{Procedure}
The study consisted of three phases, including an initial training session, the main experimental trials, and a concluding interview. 
The total session duration was approximately 45 minutes per participant.

\paragraph{Initial Setup and Training (10 mins).}
Participants were introduced to the study purpose and confirmed their consent. 
A training session was then conducted using another demo paper excerpt~\cite{bender2021dangers} to familiarize them with both the \textit{Basic} and \tool interfaces.

\paragraph{Experimental Trials (25 mins).}
Each trial began with participants reviewing a reading primer to quickly grasp the core themes of the paper. 
They then read the paper excerpt under the assigned condition, with a time limit of six minutes. We imposed this limit to simulate the focused, time-conscious nature of reviewing a paper under a deadline, though participants could proceed to the next phase once they finished reading. 
The reading time was recorded as an objective measure.
After the reading phase, participants moved on to four comprehension tasks related to the excerpt. 
The questions were presented on a printed sheet to avoid potential distractions or system effects from the digital interface. Participants were instructed to use a think-aloud protocol to verbally explain their reasoning as they navigated the text and table.
Following each condition, participants completed two questionnaires: the raw NASA-TLX to evaluate perceived cognitive workload and the System Usability Scale (SUS) to assess perceived usability. 
This procedure was repeated for both conditions, ensuring that usability and workload ratings were collected separately and immediately after each trial to minimize recall bias.
All sessions were conducted one-on-one in a quiet lab setting. 
Screen activity and audio were recorded, with participants' consent, to capture both interaction behaviors and think-aloud protocols for later analysis.

\paragraph{Semi-Structured Interview (10mins).}
At the end of the session, the experimenter conducted a semi-structured interview.
The interview provided participants with an opportunity to reflect on their experience. 
The experimenter initiated discussions about their preferred condition, the most useful aspects of the system, their information-seeking strategies, and whether they would use such a system in a real-world context.

\begin{figure*}[t!]
  \centering
  \includegraphics[width=0.99\linewidth]{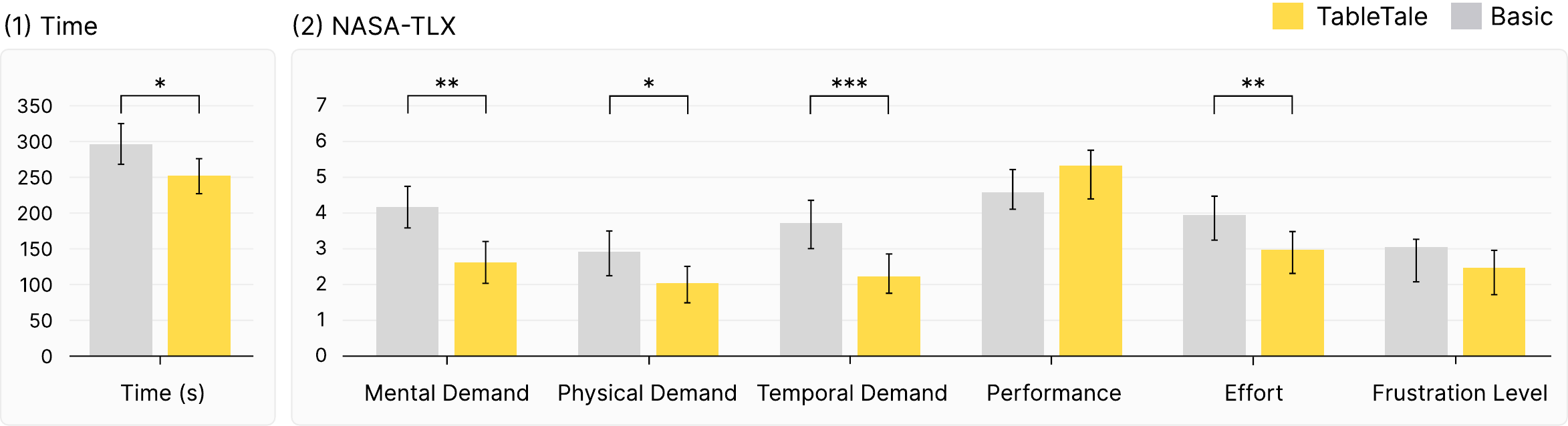}
  \caption{Quantitative results comparing the \textit{Basic} and \tool conditions. 
  (A) Task completion time was significantly reduced with \tool. 
  (B) NASA-TLX ratings showed that \tool lowered perceived workload in four dimensions (mental, physical, temporal demand, and effort), with no significant differences in performance or frustration. 
  Error bars indicate 95\% confidence intervals. 
  Wilcoxon signed-rank tests were used, with significance reported as $p < .05$ (*), $p < .01$ (**), and $p < .001$ (***).}
\Description{Two bar charts comparing Basic (gray) and TableTale (yellow) conditions. 
Panel A shows Time in seconds on a scale of 0 to 350. The Basic bar reaches nearly 300, while the TableTale bar is lower, around 250. A bracket with one asterisk indicates a significant difference. 
Panel B shows NASA-TLX ratings on a scale of 0 to 7 across six metrics. 
For Mental Demand, Basic is above 4 while TableTale is below 3 (marked with **). 
For Physical Demand, Basic is near 3 while TableTale is near 2 (marked with *). 
For Temporal Demand, Basic is near 4 while TableTale is near 2, showing the largest reduction (marked with ***). 
For Performance, Basic is above 4 while TableTale is above 5; no significance is marked. 
For Effort, Basic is near 4 while TableTale is near 3 (marked with **). 
For Frustration Level, both conditions are around 3 with no significant difference marked. 
Error bars are present on all bars.}

  \label{fig:rating}
\end{figure*}

\subsubsection{Measures and Data Analysis}

We employed three complementary measures and corresponding analyses:

\begin{enumerate}
    \item \textbf{Objective measures}: We recorded reading time, task performance, and interaction logs. 
    To assess performance accuracy, participants' verbalized reasoning was coded into three categories based on their coverage of relevant table elements: \textit{correct} (fully covers all relevant evidence), \textit{partially correct} (provides incomplete evidence), or \textit{unanswered} (no valid explanation). 
    Two independent annotators coded all responses, resolving disagreements through discussion. 

\item \textbf{Subjective measures}: We assessed usability and workload using the SUS and NASA-TLX, respectively, alongside overall satisfaction derived from interview comments. 
As the questionnaire data did not meet assumptions for parametric tests, we compared scores using non-parametric Wilcoxon signed-rank tests. 
Effect sizes are reported throughout.

    \item \textbf{Qualitative measures}: We collected audio recordings from both the think-aloud task sessions and post-study interviews. 
    These recordings were transcribed and subjected to an inductive thematic analysis. 
    Two authors performed open coding on the transcripts, followed by iterative discussions to refine codes and identify core themes regarding perceived usefulness, system drawbacks, and design suggestions.
\end{enumerate}

\subsection{Quantitative Study Results}
To answer \textbf{RQ1}, we analyzed participants’ workload ratings, usability evaluations, reading efficiency, and task performance across conditions. 
Overall, the quantitative results consistently showed that \tool assisted readers more effectively than the \textit{Basic} interface. 
Specifically, \tool significantly reduced perceived workload (NASA-TLX), achieved higher usability ratings (SUS), shortened reading time, and eliminated unanswered comprehension tasks.

\subsubsection{NASA-TLX}
We measured subjective workload using the six NASA-TLX subscales on a 7-point Likert scale.
Results indicate that our system significantly reduced perceived workload in four dimensions, see Fig.~\ref{fig:rating}(2): 
\textit{Mental Demand} \smallpar{$z = -3.800, p = .0025$}, 
\textit{Physical Demand} \smallpar{$z = -3.400, p = .0158$}, 
\textit{Temporal Demand} \smallpar{$z = -3.957, p = .0007$}, 
and \textit{Effort} \smallpar{$z = -3.514, p = .0096$}. 
No significant differences were found for \textit{Performance} and \textit{Frustration}. 
Effect sizes were consistently large across significant dimensions, suggesting that participants experienced lower cognitive and temporal pressure, as well as reduced physical effort, when using our system.

\subsubsection{System Usability Scale}
Participants rated the usability of both systems on a 7-point Likert scale. 
\tool received a higher usability score \smallpar{$M=5.65, SD=0.48$} than the \textit{Basic} condition \smallpar{$M=5.11, SD=0.76$}.
This indicates that while both interfaces were considered usable, \tool was perceived as more intuitive and supportive. 
Item-level analyses further revealed significant advantages of \tool on frequent-use intention (Q1), functional integration (Q5), consistency (Q6), ease of learning (Q8), and user confidence (Q9), underscoring its stronger potential for adoption in future academic reading workflows.

\subsubsection{Reading Time}
We observed a significant reduction in task completion time when using \tool, as shown in Fig.~\ref{fig:rating}(1). 
On average, participants spent 4:56 minutes \smallpar{$95\%$ CI = [4:28, 5:25]} under the baseline condition, 
compared to 4:12 minutes \smallpar{$95\%$ CI = [3:47, 4:36]} with \tool. 
The difference was statistically significant \smallpar{Wilcoxon $z = -2.886, p = .0119$}, 
with a large effect size \smallpar{$r = 0.589$}. 
These results suggest that \tool not only reduced subjective workload but also improved reading efficiency.

\subsubsection{Task Performance}
Across most trials, participants successfully interpreted the target sentences and located the corresponding tabular evidence. 
Because the majority of responses were correct, we focus here on the unanswered and partially correct cases. 
In the \textit{Basic} condition, we observed 17 unanswered and 3 partially correct responses. 
By contrast, in the \tool condition, only 2 responses were coded as partially correct, and no unanswered cases were observed. 
We further elaborated on the reasons behind these outcomes in our qualitative analysis (\S\ref{sssec:qualitative_point1}).


\subsection{Qualitative Feedback}

To address \textbf{RQ2}, we analyzed participants’ comments from the post-task interviews. 
The results highlight both benefits and drawbacks of \tool, including faster evidence lookup but unresolved semantic challenges, divergent preferences for multi-granularity cues, varied reading strategies, and suggestions for personalization and design improvements.

\subsubsection{\tool Accelerated Evidence Lookup but Left Deeper Semantic Understanding Unresolved}

\label{sssec:qualitative_point1}
Participants widely recognized that \tool improved efficiency in locating tabular evidence \smallpar{N=15}, effectively mitigating the burden of navigating large tables. 
Specifically, they cited benefits such as time savings \smallpar{N=7}, reduced cognitive load in information retrieval \smallpar{N=6}, and external memory support \smallpar{N=2}.
As U11 explained, \textit{``without the locator, I’d be flipping back and forth so many times…''}. 
Even when the semantic meaning was unclear, \tool helped participants identify the objects of comparison, anchoring the reference points within complex sentences.
However, these mechanical advantages did not guarantee semantic interpretation. 
While \tool clarified what was being compared, it did not explain why. 
Participants \smallpar{N=4} noted that the reasoning behind the data remained elusive: \textit{``I can figure out the numbers, but I still have to think actively to grasp the meaning''} \smallpar{U4}. 
In conclusion, while \tool provided immediate visibility of data evidence, it could not act as a substitute for explainability, leaving the final mile of semantic reasoning to the user.

\subsubsection{Progressive Disclosure Reconciled the Conflict between Navigational Guidance and Reading Flow Disruption}

Participants expressed diverse preferences regarding the granularity of visual links, driven by their varied needs for detail versus overview. 
A subset of participants \smallpar{N=4} preferred sentence-level highlights to avoid cognitive overhead, explicitly aiming to \textit{``prevent disrupting the reading flow''} \smallpar{U2}. 
Conversely, less familiar participants relied on mention-level connections as essential scaffolds to ground their understanding.
The progressive disclosure design effectively reconciled this tension. 
Observational data confirmed high engagement with this layered interaction: with the exception of three participants \smallpar{U3, U7, U18}, the vast majority consistently adopted the progressive strategy, activating sentence cues first before drilling down to specific mentions. 
Participants attributed this balance to the interaction's \textit{``modular nature''} \smallpar{U4}, noting that it felt predictable and \textit{aligned with expectations''} \smallpar{U11, U18}.
Because details were \textit{``triggered on demand''} \smallpar{U3, U13} and scoped locally, the visual cues remained helpful yet \textit{``not overwhelmingly numerous''} \smallpar{U14}.

\subsubsection{Reducing Comprehension Costs Encouraged Readers to Revisit Data-Dense Paragraphs across Diverse Reading Contexts}

Our interviews indicated that \tool partly reshaped participants’ attitudes toward data-dense sections. 
Several participants \smallpar{N=4} admitted that they typically skipped these passages in their daily academic reading, describing the experience as \textit{``boring''} \smallpar{U13} or \textit{``frustrating''} \smallpar{U11}.
With \tool’s support, however, the cost of locating relevant evidence was reduced, making them more willing to engage with such text.
For example, U8 reported that, with \tool, he was able to first conduct a quick pass and then perform close reading without feeling overloaded.
Others highlighted the importance of role differences: as researchers, they tended to prioritize methods and ideas, leaving results aside; but as reviewers, they felt obliged to examine the results carefully to judge the magnitude and reliability of improvements \smallpar{U22, U24}.  
We also observed distinct reading pathways. 
Some readers were highly number-focused, remarking that \textit{``whenever I see numbers, I want to verify their source and accuracy''} \smallpar{U6}, while others preferred to first make sense of the data claim and then validate the numbers through links \smallpar{U9}. 
In sum, \tool acted less as a rigid explainer and more as adaptable scaffolding, prompting users to revisit results while accommodating their diverse reading contexts and verification habits.

\subsubsection{Beyond Lookup: Expectations for Customization, Persistency, and Extensibility}

Participants’ feedback extended beyond the core linking mechanics in \tool, highlighting a broader demand for flexibility, ecosystem integration, and aesthetic refinement.
First, participants \smallpar{N=3} expressed a desire to shift from passive viewing to active knowledge curation, for instance, envisioning the ability to selectively retain, edit, and save specific links to \textit{``enrich personal notes''} \smallpar{U11}.
Second, there was a strong call for integration with participants’ existing reading ecosystems. Several \smallpar{N=4} noted that they hoped to use \tool as a plug-in within their current academic PDF readers. 
Others emphasized the importance of connecting \tool with commonly used utilities, such as academic translation aids \smallpar{N=6}.
Finally, aesthetic appeal and visual comfort were highlighted as prerequisites for adoption. Participants \smallpar{U4, U13} stressed that the interface should remain elegant and unobtrusive, with specific requests for environmental adaptations (e.g., night mode) \smallpar{U6} and flexible control over visual density to avoid clutter.

\section{DISCUSSION}

In this section, we synthesize our findings into design implications and potential avenues for future research.
We also reflect on current limitations regarding reliability and scalability, and suggest future directions for interactive scholarly communication.

\subsection{Design Implications}

\subsubsection{Visual Cues on Demand}  
A key finding from our study is that readers benefit from on-demand visual cues shown at different levels of detail. Readers consult tables with different intentions: sometimes they only want to quickly verify whether a high-level claim in the text holds true, while in other cases they are motivated to dig deeper, tracing evidence down to specific sentences or even single cells. 
If all possible connections are displayed simultaneously, readers can easily become overloaded and distracted, which reduces comprehension. 
By contrast, our system demonstrates the benefits of progressive disclosure, where information is revealed incrementally in response to user actions and goals. 
This strategy echoes the well-known information visualization principle of \textit{``overview first, zoom and filter, then details-on-demand''}~\cite{shneiderman2003eyes}. 
Importantly, this principle is not limited to tabular data. 
Similar strategies could be applied to other document components such as figures~\cite{visjudgeben2025xie}, formulas, or references, providing readers with more control over how much information they see at a given moment. 
Taken together, this suggests a more general and scalable method for supporting navigation and comprehension in complex documents~\cite{hullman2011visualization}.  

\subsubsection{Toward Interactive Table Narratives}  
Our approach also suggests that tables can evolve from being static blocks of numbers into active parts of a narrative. 
Rather than isolating data in stand-alone tables, we can integrate them more directly into the argumentative structure of a paper. 
By doing so, we enable smoother transitions between claims in the text and the evidence that supports them. 
This creates opportunities to build dynamic storytelling flows that not only enhance readability but also allow for richer engagement with research findings.
Beyond research articles, such flows could be adapted into new formats such as interactive summaries, explorable explanations, or even animated video that make complex studies easier to share and understand~\cite{segel2010narrative}. 
In this sense, our work connects directly with broader developments in computational storytelling and science communication. 
Interactive tables can thus be viewed as an important building block for communicating research in ways that are not only more transparent but also more engaging for diverse audiences.

\subsubsection{Position-aware Grounding for Future QA and Generative Tasks}  
Finally, our framework shows how precise, position-aware grounding of information can benefit both human readers and AI systems. 
By linking textual claims to their exact positions within a table, \tool provides readers with clear visibility into where specific evidence originates, strengthening both trust and transparency. 
At the same time, this fine-grained alignment produces valuable signals for AI systems. 
For example, it supports tasks such as table-based question answering~\cite{text2sql2025liu,deepeye2025li}, visual question answering~\cite{zhang2024mar}, automated evidence retrieval~\cite{zhang2025datamosaic,tang2024verifyai}, and retrieval-augmented generation~\cite{lewis2020retrieval}. 
Such signals allow models to reason more effectively about structured data rather than relying solely on unstructured text. 
From this perspective, our system should not only be seen as a tool that enhances the reading experience but also as a foundation for building intelligent data agents~\cite{zhu2025survey} that can reliably interpret and reason with tabular information. 
This has direct implications for both the HCI and NLP communities, highlighting opportunities for closer collaboration at the intersection of human-centered design and AI.

\subsection{Limitations}
While our results are encouraging, several practical considerations remain that highlight directions for future improvements rather than fundamental constraints of the approach.  

\paragraph{Reliability of Generative Linking.}
While our evaluation demonstrates the efficacy of agent-based multi-granular text-table linking, LLM-based approaches inherently face challenges regarding hallucinations and interpretation errors, particularly with complex table structures or long contexts. 
As noted by Kim et al.~\cite{kim2018facilitating}, false positive links may mislead readers. 
In practical settings, ensuring a high-fidelity linking schema could be achieved through human-in-the-loop verification or the integration of automated consistency-checking mechanisms.

\paragraph{Dependence on External APIs.}
Our pipeline currently leverages third-party APIs for PDF parsing and table recognition. 
While the tool is effective in many standard cases, it can occasionally introduce minor inconsistencies in bounding box detection, merged-cell interpretation, or header extraction.
Such issues may propagate into the linking stage. 
However, it is important to note that these challenges are not unique to our framework but rather reflect the broader state of the art in table recognition, which is still an active area of research. 
As APIs evolve and improve, our system can directly benefit from these advancements without requiring major architectural changes.  

\paragraph{Scalability to Complex Tables.}
Although our method handles a wide range of table structures, particularly challenging cases remain, such as very large tables spanning multiple pages or those with deeply nested headers. 
Addressing these situations may require richer table representations that can encode hierarchical relationships more effectively. 
Another promising avenue is to explore retrieval-enhanced methods or long-context LLMs, which could reason jointly over extended textual and tabular content. 
These directions represent opportunities for extending the current framework rather than inherent limitations.  

\paragraph{Domain Generalizability.}
Our analysis primarily focuses on Computer Science literature, which typically employs structured data tables with consistent formatting and stylistic conventions.
However, this scope does not fully capture the diversity of the broader scientific corpus.
In fields such as medicine, biology, or the social sciences, qualitative tables (e.g., classification or textual summaries) are common and exhibit greater structural and discursive variability. 
Adapting \tool to these domains may require further investigation and tailored adjustments to the recognition and linking modules.

\subsection{Future Work}

We outline future work across three dimensions: enhancing methodological robustness, refining adaptive interactions, and expanding into broader applications.

\paragraph{Methodological Level.}
A primary focus is on enhancing the reliability and robustness of the alignment pipeline.
To mitigate the hallucination and error risks identified in our limitations, future research should move beyond static prompting toward agentic workflows that utilize external tools for numerical verification.
Developing rigorous uncertainty estimation metrics is also crucial, as quantifying confidence levels could allow the system to suppress low-confidence links and avoid misleading readers.
Moreover, complex table structures may require dedicated alignment strategies.
Standard serialization often breaks down on large or deeply nested tables, suggesting the need for representations that more faithfully capture their spatial and hierarchical structure.

\paragraph{System Level.}
On the system design side, our findings suggest the need for deeper support of multi-granularity interaction. 
Future systems could evolve toward adaptive or personalized interaction strategies, dynamically adjusting the granularity of cues to a reader’s expertise, task, or intent. 
Moreover, interaction mechanisms such as narrative-driven table slides should be more seamlessly integrated into reading environments, enabling readers to navigate between data and text in a fluid manner. 
Such integration could also support hybrid reading modes, where readers switch flexibly between overview, detail, and cross-references.

\paragraph{Application Level.}
At the application level, the potential of table–text linking extends beyond reading assistance. 
By transforming static tables into operable and interactive objects, the approach enables tables to function as active components in a paper’s argumentation, resonating with the vision of \textit{``living papers''}~\cite{heer2023living}.
For writing support, such capabilities could help authors check reference completeness and detect inconsistencies between textual claims and tabular evidence. 
In educational contexts, the same mechanisms can scaffold novice readers' understanding of complex, data-rich articles. 
Looking forward, extending augmentation from tables to figures, equations, and other modalities may lead to a unified multimodal reading and writing ecosystem that enhances the accessibility and dissemination of scientific knowledge.

\section{Conclusion}

Data tables serve as critical instruments for communicating insights in scientific papers.
Yet their interpretation is inseparable from the surrounding narrative. 
This dependency imposes a heavy cognitive load on readers, who must constantly reconcile textual claims with tabular evidence.
In this work, we contributed three advances to address this challenge.
First, through content analysis and interviews, we established a layered account of how readers engage with data tables, identifying mention types and alignment patterns that shape comprehension. 
Second, we instantiated these insights in \tool, an augmented reading interface that leverages an LLM-agent-driven pipeline to construct a document-level linking schema and renders it through cascaded visual cues.
Third, our user study demonstrated that \tool improved efficiency, accuracy, and confidence in interpreting narratives centered on data tables.
Taken together, our efforts aim to deepen the understanding of the narrative interplay between text and data tables and point toward new directions for augmented scholarly reading.

\begin{acks}
The authors wish to thank the anonymous reviewers for their valuable comments. 
This paper was supported by the NSF of China (62402409, 72371217); the Guangzhou Industrial Informatics and Intelligence Key Laboratory No. 2024A03J0628; the Nansha Key Area Science and Technology Project No. 2023ZD003; Project No. 2021JC02X191; Guangdong Basic and Applied Basic Research Foundation (2023A1515110545); Youth S\&T Talent Support Programme of Guangdong Provincial Association for Science and Technology (SKXRC2025461); the Young Talent Support Project of Guangzhou Association for Science and Technology (QT-2025-001); Guangzhou Basic and Applied Basic Research Foundation (2025A04J3935); and Guangzhou-HKUST(GZ) Joint Funding Program (2025A03J3714)
\end{acks}

\balance
\bibliographystyle{ACM-Reference-Format}
\bibliography{reference}

\end{document}